# Behavioral and Population Data-Driven Distribution System Load Modeling

Isaac Bromley-Dulfano, Xiangqi Zhu, and Barry Mather
Power Systems Engineering Center
National Renewable Energy Laboratory
Golden, CO, USA
isaac.bromleydulfano@nrel.gov , xiangqi.zhu@nrel.gov, barry.mather@nrel.gov

*Abstract*—Distribution system residential load modeling and analysis for different geographic areas within a utility or an independent system operator territory are critical for enabling small-scale, aggregated distributed energy resources to participate in grid services under Federal Energy Regulatory Commission Order No. 2222 [1]. In this study, we develop a methodology of modeling residential load profiles in different geographic areas with a focus on human behavior impact. First, we construct a behavior-based load profile model leveraging state-of-the-art appliance models. We simulate human activity and occupancy using Markov chain Monte Carlo methods calibrated with the American Time Use Survey data set. Second, we link our model with cleaned Current Population Survey data from the U.S. Census Bureau. Finally, we populate two sets of 500 households using California and Texas census data, respectively, to perform an initial analysis of the load in different geographic areas with various group features (e.g., different income levels). To distinguish the effect of population behavior differences on aggregated load, we simulate load profiles for both sets assuming fixed physical household parameters and weather data. Analysis shows that average daily load profiles vary significantly by income and income dependency varies by locality.

## NOMENCLATURE

| | |
|---|---|
| $age$ | ATUS respondent age |
| ATUS | American Time Use Survey |
| $cat$ | ATUS respondent occupation category |
| $C_h$ | HVAC equivalent heat capacity (J/°C) |
| $C_p$ | Specific heat of water (J/kg °C) |
| $CS$ | Lighting calibration scalar |
| $i$ | Current activity state |
| $I$ | Irradiance (W/m$^2$) |
| $I_{max}$ | Irradiance lighting threshold (W/m$^2$) |
| $j$ | Transitioned activity state |
| $l$ | Lightbulb index |
| $N$ | Number of ATUS respondents |
| $occ$ | ATUS respondent occupation status |
| $O_{eff}$ | Lighting effective occupancy |
| $par$ | ATUS respondent parental status |
| $P$ | Activity transition probability |
| $Q_h$ | HVAC equivalent heat rate (W) |
| $Q_w$ | Water heater power input (W) |
| $RF$ | Lighting relative use factor |
| $R_h$ | HVAC equivalent thermal resistance (°C/W) |
| $R_w$ | Water heater tank thermal resistance (m$^2$ °C/W) |
| $SA$ | Water heater tank surface area (m$^2$) |
| $t$ | Time |
| $T_a$ | Ambient air temperature (°C) |
| $T_h$ | Water heater tank temperature (°C) |
| $T_{inc}$ | Water heater incoming water temperature (°C) |
| $T_{int}$ | Interior air temperature (°C) |
| $V$ | Water heater tank volume (m$^3$) |
| $W_D$ | Hot water demand (L/s) |
| $\Delta t$ | Time step |

## I. INTRODUCTION

To address the ongoing climate crisis, major shifts are taking place in both the bulk power and distribution systems. Investments in utility-scale renewable energy continue to soar [2], and distributed energy resources (DER) are increasingly considered essential for decarbonized grid operations [3]. In 2020, the Federal Energy Regulatory Commission (FERC) released Order No. 2222 [1], which enables small-scale, aggregated DERs to engage in grid services. One small-scale DER service of interest is demand response, in which aggregators reduce or shift flexible residential loads during peak load hours, typically through consumer incentives or direct appliance control. With the expansion of demand response programs under FERC 2222, planners require advanced load profile models and analysis to predict outcomes for individual residential consumers and system reliability at large [4]–[5].

Existing literature has taken several approaches to modeling loads in distribution systems, particularly residential household loads, with both "top-down" [6]–[8] and "bottom-up" [9]–[19] methods. Although top-down methods require relatively simple data inputs, they typically do not distinguish load contributions of individual components. This is undesirable, especially under FERC 2222. Without the knowledge of small group or individual load profiles, it is difficult to predict and evaluate the performance of demand response programs.

Conversely, bottom-up studies model subcomponents within a complex load and aggregate to obtain the expected demand profile. To provide visibility to load aggregators and grid operators for demand response potential analysis, we focus on the bottom-up approach in this paper.

The state of the art offers a range of modeling options (e.g., different physical appliance models and population clustering methods), upon which we can build our model. A popular bottom-up strategy in the state of the art [9]–[19] is to model human activity using Markov Chain processes calibrated to behavioral time use surveys (TUS) and derive the associated residential load profiles. These studies associate each simulated human activity (cooking, laundry, etc.) with residential appliances (oven, washer, dryer, etc.). They use physical models or measured data to represent the load profiles of active appliances. Studies have applied this method using public TUS data in the United Kingdom [9], [12], [18], [20], Sweden [10], [16], France [21], the Netherlands [17], and the United States [11], [13], [22], [23]; and several have validated their behavioral load models against metered household data [9]–[11], [17], [19].

Although behavioral load modeling is growing in the realm of research, only a limited number of studies have considered behavior-defined load modeling across geographic and socioeconomic dimensions. Diao et al. [22] take a necessary first step in clustering population activity patterns, but the question remains how these activity patterns—and their resulting load profiles—will aggregate for distinct geographically distributed populations. For example, to gauge demand response potential, aggregators need to know how load profiles differ among small areas within a larger city. If available census data exist to capture population distributions within the city, a behavior-based model can provide insights that would otherwise require detailed metered data.

To address this question, we propose linking U.S. census data with a behavior-based household load model to capture geographic variations in load profiles at the aggregator level or census block level. This paper presents a proof of concept. First, we construct a behavior-based household load profile model by integrating the model approaches in the literature. Second, by using a generic set of appliances and fixed weather data, we model the load profiles for two sets of 500 households populated with occupants from the California and Texas census subsets, respectively. By fixing the household characteristics and weather profiles, we demonstrate the sensitivity of the aggregated household load profiles to the population characteristics and associated behavior pattern differences.

The rest of the paper is organized as follows: Section II introduces the methodology for behavior-based load profile modeling. Section III presents an initial analysis of the aggregated load profile for a representative socioeconomic dimension—income level. Section IV concludes the paper and discusses potential future work.

## II. MODELING METHODOLOGY

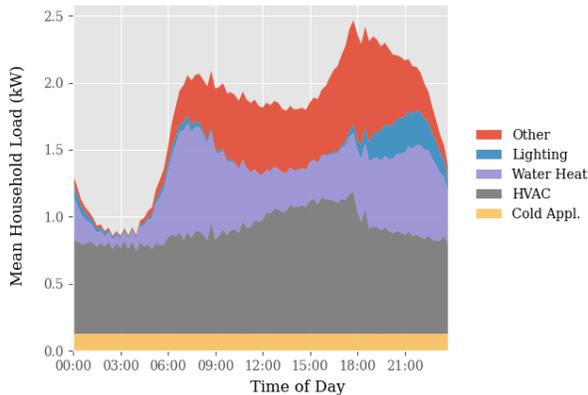

Fig. 1 Mean household load profile sample.

We construct a household load profile model following three main steps: (1) We model occupancy for household members based on their personal characteristics, i.e., age, work status, work type, and parental status; (2) We translate household member activities into appliance-use profiles, such as hot water consumption, heating, ventilating, and air-conditioning (HVAC) needs, etc.; and (3) We use appliance load profile models or power conversion factors in conjunction with the output from Step 2 to derive the final load profile.

The model outputs minute-resolution load profiles by end use (a sample is shown in Fig. 1). To connect the model with U.S. census data, we directly populate households with members from the 2019 Current Population Survey (CPS).

### A. Occupancy Modeling

To model the activity and occupancy of household members, we use a Markov chain Monte Carlo simulation calibrated with 2016–2019 American Time Use Survey (ATUS) data [24]. The ATUS contains detailed, 24-hour activity logs for 50,000 respondents on both weekdays and weekends. To clean the ATUS data, we classify each activity into one of nine activity states, as shown in Table. I. Additionally, we label each respondent by their age, work status, work type, and parental status because these indicators have been shown to affect activity patterns [25], [26]. Our model follows the literature [11], [16], [22], [23], [26] and uses hourly transition probability matrices for the weekday and weekend to sample transitions between activities. We populate the weekday or weekend transition probability matrices according to Equation (1), which describes the probability of an arbitrary household member transitioning between two activities at a given time. Finally, we sample transitions between activities every 10 minutes.

$$P_{age,occ,cat,par}^{t,i,j} = \frac{N_{age,occ,cat,par}^{t,i,j}}{N_{age,occ,cat,par}^{t,i}} \quad (1)$$

Table I. Classified ATUS Activity States.

| # | Activity State | # | Activity State |
|---|---|---|---|
| 0 | Away | 5 | Cleaning |
| 1 | Sleeping | 6 | Laundry |
| 2 | Grooming | 7 | Leisure |
| 3 | Cooking | 8 | Other |
| 4 | Dishwashing | - | - |

### B. Appliance Modeling

#### 1) HVAC

To simulate HVAC load profiles, we use a thermal model described by Lu [27]. The model characterizes interior temperature as a function of ambient temperature and three effective parameters tuned to the behavior of a typical household (Equation (2)).

$$T_{int}^{t+1} = T_a^{t+1} + Q_h R_h - (T_a^{t+1} + Q_h R_h - T_{int}^t) \cdot e^{-\frac{\Delta t}{R_h C_h}} \quad (2)$$

The time variant equivalent heat rate, $Q_h$, indicates the active HVAC state (i.e., heating, cooling, neither). We select the HVAC state to maintain interior temperatures within the set point deadband. Although this simplified model does not capture differences between households (i.e., physical house size, level of insulation), it is well suited for our central experiment, which uses a fixed household for the entire population (Section III). To integrate occupancy in the HVAC model, we adjust the set point when all household members are away.

#### 2) Water Heater

To account for water heater loads, we use a thermal model presented by Dolan et al. [28]. The model characterizes tank temperature behavior as a function of ambient temperature, tank size, incoming water temperature, and hot water demand (Equations (3)–(7)).

$$T_h^{t+1} = T_h^t \cdot e^{-\frac{\Delta t}{R'C_w}} + [G \cdot T_a + B^t \cdot T_{in} + Q_w] \cdot R' \cdot [1 - e^{-\frac{\Delta t}{R'C_w}}] \quad (3)$$

$$R' = \frac{1}{G + B^t} \quad (4)$$

$$G = \frac{SA}{R_w} \quad (5)$$

$$B = W_D^t * C_p \quad (6)$$

$$C_w = C_p * V \quad (7)$$

To integrate human behavior in the water heater model, we sample hot water consumption events with predetermined consumption profiles associated with occupant activity states (Table. II).

Table. II: Hot Water Consumption Profiles.

| | Activity State | Hot Water Consumption Activity | Flow Rate (L/min.) | Duration (min.) |
|---|---|---|---|---|
| 0 | Away | - | - | - |
| 1 | Sleeping | - | - | - |
| 2 | Grooming | Showering | 8.0 | 8 |
| 3 | Cooking | Washing hands, filling pots | .1 | 10 |
| 4 | Dishwashing | Filling washer | 4.0 | 5 |
| 5 | Cleaning | Mopping, washing counters | 1.2 | 5 |
| 6 | Laundry | Filling washer | 2.5 | 30 |
| 7 | Leisure | - | - | - |
| 8 | Other | - | - | - |

*3) Lighting*

To model lighting loads, we use an agent-based method presented by Richardson et al. [29]. Following their method, we sequentially sample turn-on events for individual lightbulbs based on outdoor global irradiance and household occupancy.

We convert absolute occupancy into effective occupancy to account for shared-use lighting. Additionally, we assign relative use factors to individual lightbulbs and include a calibration scalar given in the literature [29], [30]. In each time step, we use Equation (8) to find the probability of turning on each lightbulb. Following turn-on events, we sample the duration for which lights remain active from a distribution provided by Richardson et al. [29].

$$P_{on,l}^t = (I^t > I_{max}) \cdot O_{eff}^t \cdot RF_l \cdot CS \quad (8)$$

*4) Cold Appliances*

To account for cold appliance loads—namely, refrigerators and freezers—we follow a simple procedure described by Muratori et al. [11]. For each appliance, we sample operation in 10-minute intervals assuming a Bernoulli distribution. The exact distribution is calculated such that the expected annual cold appliance consumption is consistent with that reported in the Residential Energy Consumption Survey [31].

*5) Other Appliances*

For all other appliances—including ovens, televisions, computers, etc.—we use power conversion factors and predetermined appliance load profiles [10], [11]. When at least one occupant is engaged in an activity state, we include the power conversion factor associated with that state in the load profile. For laundry and dishwashing events, we assume a predetermined load profile because the associated appliances typically remain active after the occupant has transitioned to another state [10]. The conversion factors and predetermined load profiles are shown in Table III.

Table. III: Power Conversion Factors and Profiles.

| | Activity State | Appliances | Load (W) | Duration (min.) |
|---|---|---|---|---|
| 0 | Away | - | - | - |
| 1 | Sleeping | - | - | - |
| 2 | Grooming | Hair dryer | 100 | - |
| 3 | Cooking | Oven, toaster, microwave | 3500 | - |
| 4 | Dishwashing | Dishwasher | 1800 | 60 |
| 5 | Cleaning | Vacuum cleaner | 1500 | - |
| 6 | Laundry | Washing machine, dryer | 425/3400 | 30/90 |
| 7 | Leisure | Television, computer | 120 | - |
| 8 | Other | - | - | - |

### III. LOAD PROFILE ANALYSIS

To demonstrate the benefit of linking census data with behavior-based load modeling at the aggregator level, we conduct an analysis to isolate the effect of population behavior on average and aggregated load profiles. First, we build a generic fixed household. We assign the model parameters in Table IV, and we use 2019 temperature and irradiance data from Austin, Texas [31]. Next, we populate two sets of 500 households using 2019 CPS census data. Because the household characteristics and weather data are fixed, the only difference between households is the number of occupants and their respective labels (i.e., age, work status, work type, parental status). For each respective set, we assign the number of occupants and occupant labels from random households in the Texas and California CPS subsets.

In our analysis, we compare the overall load profiles and load profiles by income for the two populations. Finally, because the California and Texas populations differ in household size distributions, we repeat the income analysis for households with only two members.

Table IV: Assumed Household Parameters.

| Parameter | Value | Units | Model(s) |
|---|---|---|---|
| $C_h$ | 40,000 | J/°C | HVAC |
| $R_h$ | .18 | °C/W | HVAC |
| $Q_h$ (heating) | 450 | W | HVAC |
| $Q_h$ (cooling) | -150 | W | HVAC |
| HVAC set point (home) | 21 | °C | HVAC |
| HVAC set point (away) | 21 ± 5 | °C | HVAC |
| HVAC deadband | 2 | °C | HVAC |
| Heater rating | 6000 | W | HVAC |
| AC rating | 4500 | W | HVAC |
| WH set point | 55 | °C | Water heater |
| WH deadband | 4 | °C | Water heater |
| $T_{inc}$ | 10 | °C | Water heater |
| $V$ | 190 | L | Water heater |
| $SA$ | 2 | m² | Water heater |
| $C_p$ | 4186 | J/kg °C | Water heater |
| $R_w$ | 1.2 | m² °C / W | Water heater |
| $Q_w$ | 3000 | W | Water heater |

| Lightbulbs | 30 | Quantity | Lighting |
| --- | --- | --- | --- |
| CS | .008 | - | Lighting |
| Refrigerator rating | 200 | W | Cold appliances |
| Freezer rating | 50 | W | Cold appliances |

*A. Aggregated Household Load*

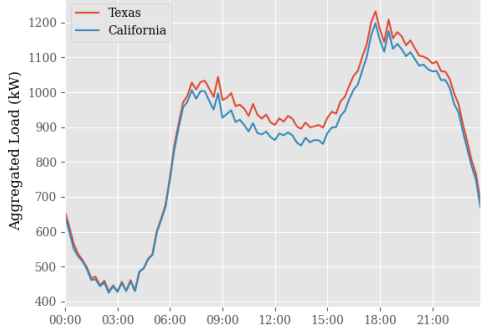

Fig. 2: Aggregated load for 500 Texas (red) and California (blue) households.

The results from the 500-household analysis indicate that the purely behavior-driven aggregated load profiles of the Texas and California samples are similar in shape and comparable in magnitude. The shape is primarily driven by the HVAC load profile (as shown in Fig. 1 in Section II); however, because the weather data are fixed, differences in HVAC consumption caused by occupancy are difficult to identify in the aggregate load profiles. Overall, the Texas sample consumes 3% more electricity annually and demands up to 6% more power during the middle hours of the day (Fig. 2). We attribute this difference to a larger average household size (2.6 occupants) relative to the California sample (2.4 occupants).

*B. Load Profiles by Income*

Households with similar characteristics, such as income, are likely to cluster in the same geographic areas; therefore, distinguishing load profiles by income bracket can capture the load differences among geographic areas that aggregators might observe within a city. Hence, we further investigate the load profile differences between income brackets.

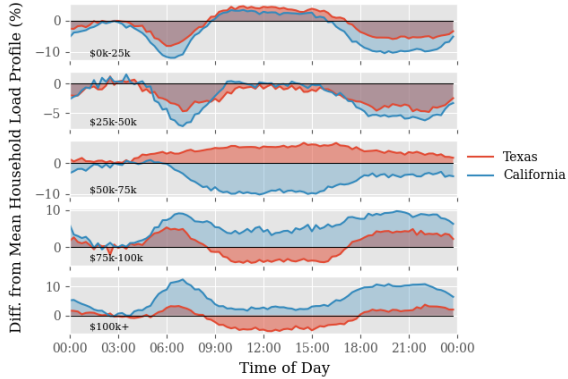

Fig. 3: Difference from mean household load profile by income bracket.

Fig. 3 shows how the mean load profile for each income bracket differs from the overall mean load profile throughout the day by state. In California, loads in the lowest income bracket are up to 11% less than the mean California loads during peak load hours. Conversely, loads in the highest income bracket are up to 12% greater than the mean loads during peak load hours. For households making $50K–75K, the Texas load profile is up to 7% greater than the mean Texas loads, whereas the California load profile is up to 10% less than the mean California loads.

We largely attribute differences between states to their respective income-dependent household size distributions. Generally, average household size grows with income, however, the largest average household size occurs in the $50K–75K bracket and the +$100K bracket for the Texas and California samples, respectively.

*C. Load Profiles by Income with Fixed Household Size*

To remove the effect of house size on load profiles, we repeat the analysis in Part III.B, but we only include census households with two occupants. Fig. 4 shows that the fixed-size California and Texas samples yield generally consistent results across incomes. Loads in lower-income households are up to 6% less than mean loads during peak residential load periods and up to 10% more during the middle hours of the day. Loads in high-income households are up to 5% greater than mean loads during the morning peak and up to 10% less during the middle hours of the day. To be clear, we can attribute the results in Fig. 4 purely to behavioral differences across income brackets.

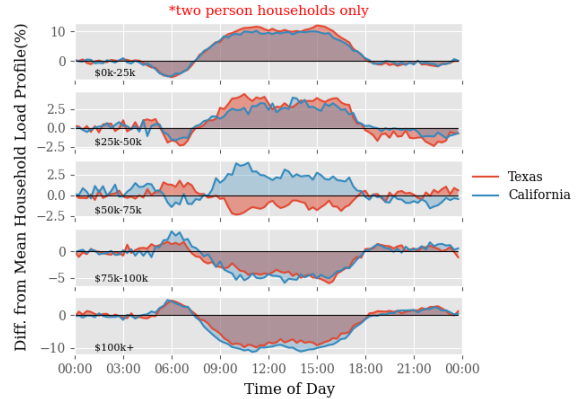

Fig. 4: Difference from mean household load profile by income bracket two-person households.

This analysis can indicate the demand response potential of small areas (i.e., neighborhood or census block level) for which there are available Census data. For example, Fig. 4 indicates that *considering only family size and behavior*, the $75K–100K households show the most potential for load shifting in the mid-to-late afternoon hours in California, whereas in Texas the $50–75K households show more potential during the same period.

IV. CONCLUSIONS AND FUTURE WORK

This paper demonstrates the potential of distinguishing load profiles and demand response potential for different geographic regions using behavior-based load models in conjunction with aggregated population characteristics. Based on previous reported work, we construct an integrated behavior-based load profile model. Then we link data from the U.S. Census Bureau CPS to the load profile model, and we compare the load profiles for two sets of 500 households populated by Texas and California census data, respectively, assuming a fixed set of appliances and weather data. From our analysis, we find that the mean load profiles differ significantly according to income in both California and Texas, which can be used by aggregators to estimate demand response potential in geographically distributed income clusters. In our future work, we will perform further analysis from more socioeconomic dimensions for different geographic areas.

ACKNOWLEDGMENTS

This work was authored by the National Renewable Energy Laboratory, operated by Alliance for Sustainable Energy, LLC, for the U.S. Department of Energy (DOE) under Contract No. DE-AC36-08GO28308. Funding provided by U.S. Department of Energy SULI program. The views expressed in the article do not necessarily represent the views of the DOE or the U.S. Government. The U.S. Government retains and the publisher, by accepting the article for publication, acknowledges that the U.S. Government retains a nonexclusive, paid-up, irrevocable, worldwide license to publish or reproduce the published form of this work, or allow others to do so, for U.S. Government purposes.

REFERENCES

[1] Federal Energy Regulatory Commission, Order No. 2222. 2020.
[2] L. Perea, Austin (Wood Mackenzie et al., "Solar Market Insight Report - 2018 Q2 - Executive Summary," 2018.
[3] P. de Martini and L. Kristov, "Distribution Systems in a High Distributed Energy Resources Future," 2015.
[4] M. Mohammed, A. Abdulkarim, A. S. Abubakar, A. B. Kunya, and Y. Jibril, "Load modeling techniques in distribution networks: a review," Journal of Applied Materials and Technology, vol. 1, no. 2, 2020, doi: 10.31258/jamt.1.2.63-70.
[5] A. J. Collin, G. Tsagarakis, A. E. Kiprakis, and S. McLaughlin, "Development of low-voltage load models for the residential load sector," IEEE Transactions on Power Systems, vol. 29, no. 5, 2014, doi: 10.1109/TPWRS.2014.2301949.
[6] E. Hirst, "A model of residential energy use," Simulation, vol. 30, no. 3, 1978, doi: 10.1177/003754977803000301.
[7] M. Saviozzi, S. Massucco, and F. Silvestro, "Implementation of advanced functionalities for Distribution Management Systems: Load forecasting and modeling through Artificial Neural Networks ensembles," Electric Power Systems Research, vol. 167, 2019, doi: 10.1016/j.epsr.2018.10.036.
[8] X. Zhu and B. Mather, "Data-Driven Distribution System Load Modeling for Quasi-Static Time-Series Simulation," IEEE Transactions on Smart Grid, vol. 11, no. 2, 2020, doi: 10.1109/TSG.2019.2940084.
[9] I. Richardson, M. Thomson, D. Infield, and C. Clifford, "Domestic electricity use: A high-resolution energy demand model," Energy and Buildings, vol. 42, no. 10, 2010, doi: 10.1016/j.enbuild.2010.05.023.
[10] J. Widén and E. Wäckelgård, "A high-resolution stochastic model of domestic activity patterns and electricity demand," Applied Energy, vol. 87, no. 6, 2010, doi: 10.1016/j.apenergy.2009.11.006.
[11] M. Muratori, M. C. Roberts, R. Sioshansi, V. Marano, and G. Rizzoni, "A highly resolved modeling technique to simulate residential power demand," Applied Energy, vol. 107, 2013, doi: 10.1016/j.apenergy.2013.02.057.
[12] E. Lampaditou and M. Leach, "Evaluating participation of residential customers in demand response programs in the UK," eceee 2005 summer study - What works & who delivers?, 2005.
[13] J. Zhu, Q. Liao, Y. Lin, W. Lei, and R. Cui, "Residential high-resolution electricity demand optimization with a cooperative PSO algorithm," Energy Reports, vol. 7, 2021, doi: 10.1016/j.egyr.2021.02.031.
[14] A. Mammoli, M. Robinson, V. Ayon, M. Martínez-Ramón, C. fei Chen, and J. M. Abreu, "A behavior-centered framework for real-time control and load-shedding using aggregated residential energy resources in distribution microgrids," Energy and Buildings, vol. 198, 2019, doi: 10.1016/j.enbuild.2019.06.021.
[15] S. Ge, J. Li, H. Liu, X. Liu, Y. Wang, and H. Zhou, "Domestic energy consumption modeling per physical characteristics and behavioral factors," in Energy Procedia, 2019, vol. 158. doi: 10.1016/j.egypro.2019.01.399.
[16] J. Widén, A. Molin, and K. Ellegård, "Models of domestic occupancy, activities and energy use based on time-use data: Deterministic and stochastic approaches with application to various building-related simulations," Journal of Building Performance Simulation, vol. 5, no. 1, 2012, doi: 10.1080/19401493.2010.532569.
[17] M. Nijhuis, M. Gibescu, and J. F. G. Cobben, "Bottom-up Markov Chain Monte Carlo approach for scenario based residential load modelling with publicly available data," Energy and Buildings, vol. 112, 2016, doi: 10.1016/j.enbuild.2015.12.004.
[18] G. Tsagarakis, A. J. Collin, and A. E. Kiprakis, "Modelling the electrical loads of UK residential energy users," 2012. doi: 10.1109/UPEC.2012.6398593.
[19] K. McKenna and A. Keane, "Residential Load Modeling of Price-Based Demand Response for Network Impact Studies," IEEE Transactions on Smart Grid, vol. 7, no. 5, 2016, doi: 10.1109/TSG.2015.2437451.
[20] I. Richardson, M. Thomson, and D. Infield, "A high-resolution domestic building occupancy model for energy demand simulations," Energy and Buildings, vol. 40, no. 8, 2008, doi: 10.1016/j.enbuild.2008.02.006.
[21] U. Wilke, F. Haldi, J. L. Scartezzini, and D. Robinson, "A bottom-up stochastic model to predict building occupants' time-dependent activities," Building and Environment, vol. 60, 2013, doi: 10.1016/j.buildenv.2012.10.021.
[22] L. Diao, Y. Sun, Z. Chen, and J. Chen, "Modeling energy consumption in residential buildings: A bottom-up analysis based on occupant behavior pattern clustering and stochastic simulation," Energy and Buildings, vol. 147, 2017, doi: 10.1016/j.enbuild.2017.04.072.
[23] Y. S. Chiou, K. M. Carley, C. I. Davidson, and M. P. Johnson, "A high spatial resolution residential energy model based on American Time Use Survey data and the bootstrap sampling method," Energy and Buildings, vol. 43, no. 12, 2011, doi: 10.1016/j.enbuild.2011.09.020.
[24] U.S. Bureau of Labor Statistics, "American time use survey," Washington, D.C., 2019.
[25] D. S. Hamermesh, H. Frazis, and J. Stewart, "Data watch the American time use survey," Journal of Economic Perspectives, vol. 19, no. 1, 2005, doi: 10.1257/0895330053148029.
[26] F. Farzan, M. A. Jafari, J. Gong, F. Farzan, and A. Stryker, "A multi-scale adaptive model of residential energy demand," Applied Energy, vol. 150, 2015, doi: 10.1016/j.apenergy.2015.04.008.
[27] N. Lu, "An evaluation of the HVAC load potential for providing load balancing service," IEEE Transactions on Smart Grid, vol. 3, no. 3, 2012, doi: 10.1109/TSG.2012.2183649.
[28] P. S. Dolan, M. H. Nehrir, and V. Gerez, "Development of a Monte Carlo based aggregate model for residential electric water heater loads," Electric Power Systems Research, vol. 36, no. 1, 1996, doi: 10.1016/0378-7796(95)01011-4.
[29] I. Richardson, M. Thomson, D. Infield, and A. Delahunty, "Domestic lighting: A high-resolution energy demand model," Energy and Buildings, vol. 41, no. 7, 2009, doi: 10.1016/j.enbuild.2009.02.010.
[30] E. J. Palacios-Garcia, A. Chen, I. Santiago, F. J. Bellido-Outeiriño, J. M. Flores-Arias, and A. Moreno-Munoz, "Stochastic model for lighting's electricity consumption in the residential sector. Impact of energy saving actions," Energy and Buildings, vol. 89, 2015, doi: 10.1016/j.enbuild.2014.12.028.
[31] U.S. Energy Information Administration, "Residential Energy Consumption Survey (RECS)," 2015.